\documentclass{pnastwo}
\usepackage{amssymb,amsfonts,amsmath}
\usepackage{graphicx}
\usepackage{graphics}
\usepackage{subfig}		
\usepackage{dcolumn}	

\contributor{Submitted to Proceedings of the National Academy of Sciences of the United States of America}
\url{www.pnas.org/cgi/doi/10.1073/pnas.0709640104}
\copyrightyear{2008}
\issuedate{Issue Date}
\volume{Volume}
\issuenumber{Issue Number}

\begin{document}
\title{The Role of Design Complexity in Technology Improvement}
\author{James McNerney\affil{1}{Santa Fe Institute, 1399 Hyde Park Road, Santa Fe, NM 87501, USA}\affil{2}{Department of Physics, Boston University, Boston, MA 02215, USA},
J. Doyne Farmer\affil{1}{}\affil{3}{LUISS Guido Carli, Viale Pola 12, 00198, Roma, Italy},
Sid Redner\affil{1}{}\affil{2}{}\affil{4}{Center for Polymer Studies, Boston University, Boston, MA 02215, USA},
\and
Jessika E Trancik\affil{1}{}\affil{5}{Earth Institute, Columbia University, New York, NY 10027}}
\maketitle

\begin{article}
\begin{abstract}

  We study a simple model for the evolution of the cost (or more generally
  the performance) of a technology or production process.  The technology can
  be decomposed into $n$ components, each of which interacts with a cluster
  of $d-1$ other, dependent components.  Innovation occurs through a series of
  trial-and-error events, each of which consists of randomly changing the
  cost of each component in a cluster, and accepting the changes only if the
  total cost of the entire cluster is lowered.  We show that the relationship
  between the cost of the whole technology and the number of innovation
  attempts is asymptotically a power law, matching the functional form often
  observed for empirical data.  The exponent $\alpha$ of the power law
  depends on the intrinsic difficulty of finding better components, and on
  what we term the {\it design complexity}: The more complex the design, the
  slower the rate of improvement.  Letting $d$ as defined above be the
  connectivity, in the special case in which the connectivity is constant,
  the design complexity is simply the connectivity.  When the connectivity
  varies, bottlenecks can arise in which a few components limit progress.  In
  this case the design complexity is more complicated, depending on the
  details of the design.  The number of bottlenecks also determines whether
  progress is steady, or whether there are periods of stasis punctuated by
  occasional large changes.  Our model connects the engineering properties of
  a design to historical studies of technology improvement.

\end{abstract}

\dropcap{T}he relation between a technology's cost $c$ and the cumulative amount produced $y$ is often empirically observed to be a power law of the form
\begin{align}
c(y) \propto y^{-\alpha}, 
\end{align}
where the exponent $\alpha$ characterizes the rate of improvement.  This rate is commonly termed the \emph{progress ratio} $2^{-\alpha}$, which is the factor by which costs decrease with each doubling of cumulative production.  A typical reported value \cite{Dutton84} is 0.8 (corresponding to $\alpha \approx .32$), which implies that the cost of the 200th item is 80\% that of the 100th item.  Power laws have been observed, or at least assumed to hold, for a wide variety of technologies \cite{Argote90,McDonald01,Dutton84}, although other functional forms have also been suggested and in some cases provide plausible fits to the data\footnote{
 Koh and Magee \cite{Koh06,Koh08}
  claim an exponential function of time (Moore's law) predicts the performance of several different technologies.  Goddard \cite{Goddard82} claims costs follow a power law in production rate rather than 
  cumulative production. Multivariate forms involving combinations of production rate, cumulative 
  production, or time have been examined by Sinclair et al. \cite{Sinclair00} and Nordhaus \cite{Nordhaus09}.}.
We give examples of historical performance curves for several different technologies in Fig. 1.

The relationship between cost and cumulative production goes under several different names, including the ``experience curve", the ``learning curve" or the ``progress function". The terms are used interchangeably by some, while others assign distinct meanings \cite{Dutton84,Thompson08}.   We use the general term \emph{performance curve} to denote a plot of any performance measure (such as cost) against any experience measure (such as cumulative production), regardless of the context.  Performance curve studies first appeared in the 19th century \cite{Ebbinghaus1885, Bryan1899}, but their application to manufacturing and technology  originates from the 1936 study by Wright on aircraft production costs \cite{Wright36}.  The large literature on this subject spans engineering \cite{Ostwald79}, economics \cite{Arrow62,Thompson08}, management science \cite{Dutton84}, and public policy \cite{IEA00}. Performance curves have been constructed for individuals, production processes, firms, and industries \cite{Dutton84}.

The power law assumption has been used by firm managers \cite{BCG72} and government policy makers \cite{IEA00} to forecast how costs will drop with cumulative production.  However, the potential for exploiting performance curves has so far not been fully realized, in part because there is no theory explaining the observed empirical relationships.   Why do performance curves tend to look like power laws, as opposed to some other functional form?  What factors determine the exponent $\alpha$, which governs the long-term rate of improvement?  Why are some performance curves steady and others erratic? By suggesting answers to these questions, the theory we develop here can potentially be used to guide investment policy for technological change.
\begin{figure}[b]
\noindent \includegraphics[width=0.48\textwidth]{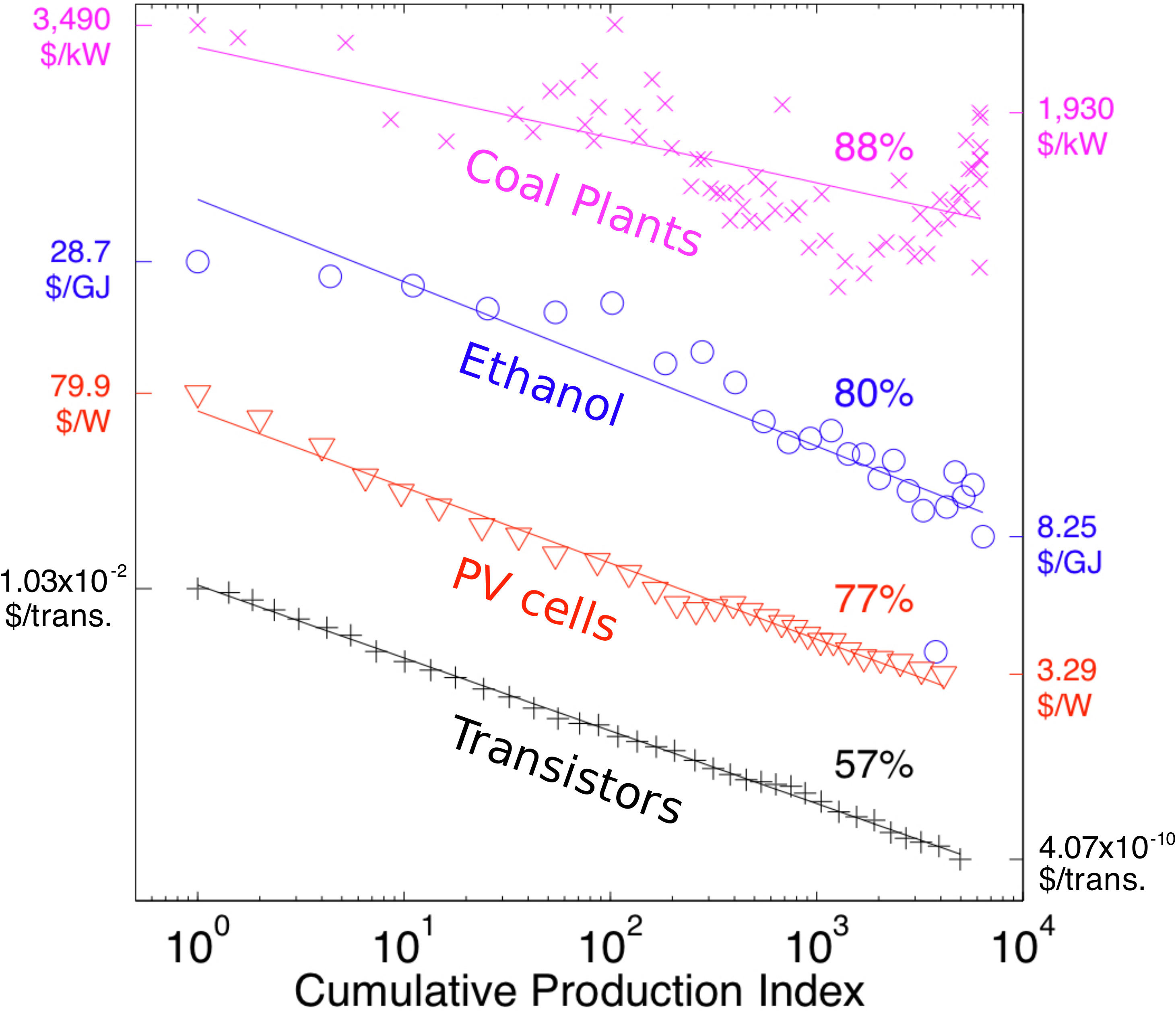}
\caption{Four empirical performance curves. Each curve was rescaled and shifted to aid comparison with a power law. The $x$- and $y$- coordinates of each series $i$ were transformed via $\log x \rightarrow a_i + b_i \log x$, $\log y \rightarrow c_i + d_i \log y$. The constants  $a_i$, $b_i$, $c_i$, and $d_i$ were chosen to yield series with approximately the same slope and range, and are given in the Supporting Information. Tick marks and labels on the left vertical axis show the first price (in real 2000 dollars) of the corresponding time series, and those of the right vertical axis show the last price. Lines are least-squares fits to a power law. Percentages are the progress ratios of the fitted power laws. Source: coal plants \cite{McNerney09}, ethanol \cite{Alberth08}, photovoltaic cells \cite{Nemet06,StratUnlim04,Maycock02}, transistors \cite{Moore65}.}
\label{performance_curves}
\end{figure}

A good example of the possible usefulness of such a theory is climate change mitigation.  Good forecasts of future costs of low-carbon energy technologies could help guide research and development funding and climate policy.  
Our theory suggests that based on the design of a technology we might be able to better forecast its rate of improvement, and therefore make better investments and better estimates of the cost of achieving low-carbon energy generation.

There have been several previous attempts to construct theories to explain the functional form of performance curves \cite{Muth86,Auerswald00,Huberman01}.  Muth constructed a model of a single-component technology in which innovation happens by proposing new designs at random.  Using extreme value theory he derived conditions under which the rate of improvement is a power law.  An extension to multiple components, called the \emph{production recipe model}, was proposed by Auerswald et al. \cite{Auerswald00}.  In their model each component interacts with other components, and if a given component is replaced, it affects the cost of the components with which it interacts.
They simulated their model and found that under some circumstances the performance curves appeared to be power laws. 

We simplify the production recipe model in order to make it more tractable.
This simplification allows us to derive the emergence of a power law, and in
particular, to derive its exponent $\alpha$.  We obtain the result that the
asymptotic rate of improvement is independent of the total number of
components $n$.  Instead we find that $\alpha = 1/(\gamma d^*)$, where
$\gamma$ measures the intrinsic difficulty of finding better components and
$d^*$ is what we call the {\it design complexity}.  When the connectivity of
the components is constant, the design complexity is equal to the
connectivity, but when connectivity varies, it can also depend on the
detailed properties of the design.  We also show that when costs are spread
uniformly across a large number of components, the whole technology undergoes
steady improvement.  In contrast, when costs are dominated by a few
components, it undergoes erratic improvement.  Our theory thus potentially
gives insight into how to design a technology so that it will improve more
rapidly and more steadily.

\section{The Model}
The production design consists of $n$ components, which can be thought of as
the parts of a technology or the steps in an industrial process.\footnote{
  The original production recipe model contained 6 parameters. The following
  simplifications were made to produce the model in this paper: Length of
  production run $T \rightarrow \infty$. Output-per-attempted-recipe-change
  $\hat{B} \rightarrow 1$. Available implementations per component $s
  \rightarrow \infty$. Search distance $\delta \rightarrow 1$.}
Each component $i$ has a cost $c_i$.  The total cost $c$ of the design is the
sum of the component costs: $c = c_1 + c_2 + \cdots + c_n$.  A component's
cost changes as new implementations for the component are found. For example,
a component representing the step ``move a box across a room'' may initially
be implemented by a forklift, which could later be replaced by a conveyor
belt.  Cost reductions occur through repeated changes to one or more
components.

Components are not isolated from one another, but rather interact as parts of
the overall design.  Thus changing one component not only affects its cost,
but also the costs of other {\em dependent\/} components.  Components may be
viewed as nodes in a directed network, with links from each component to
those that depend on it.  The relationship between the nodes and links can
alternately be characterized by an adjacency matrix.  In systems engineering
and management science this matrix is known as the \emph{design structure
  matrix} or DSM \cite{Steward81,Eppinger94,Baldwin00}.  A DSM is an $n
\times n$ matrix with an entry in row $i$ and column $j$ if a change in
component $j$, the modifying component, affects component $i$, the dependent
component (Fig. 2).  The matrix is usually binary
\cite{Whitney99,Rivkin07}; however, weighted interactions have also been
considered \cite{Chen07}.  DSMs have been found to be useful in understanding
and improving complex manufacturing and technology development processes.

\begin{figure}[t]
\noindent \includegraphics[width=0.24\textwidth]{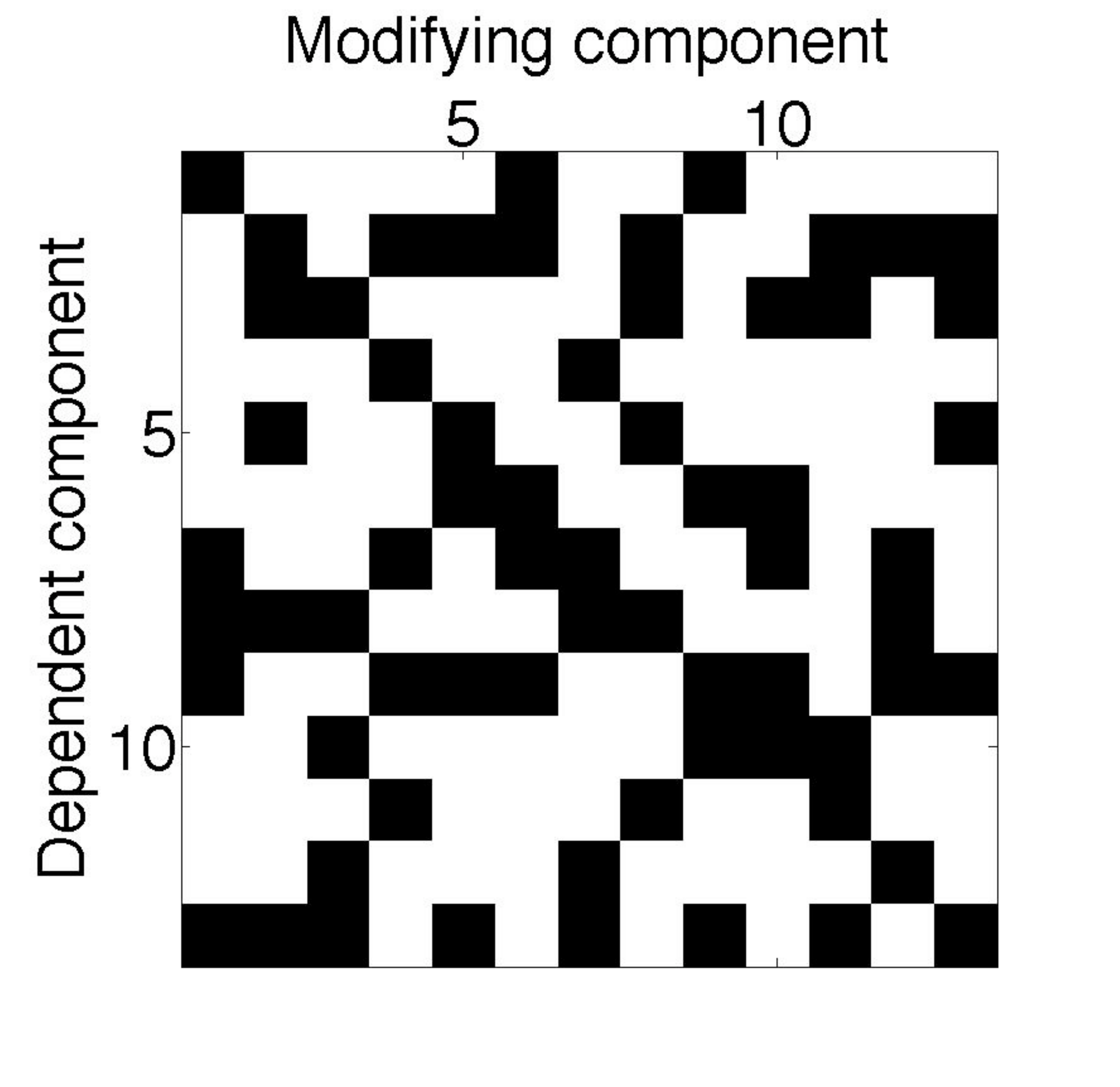}\includegraphics[width=0.24\textwidth]{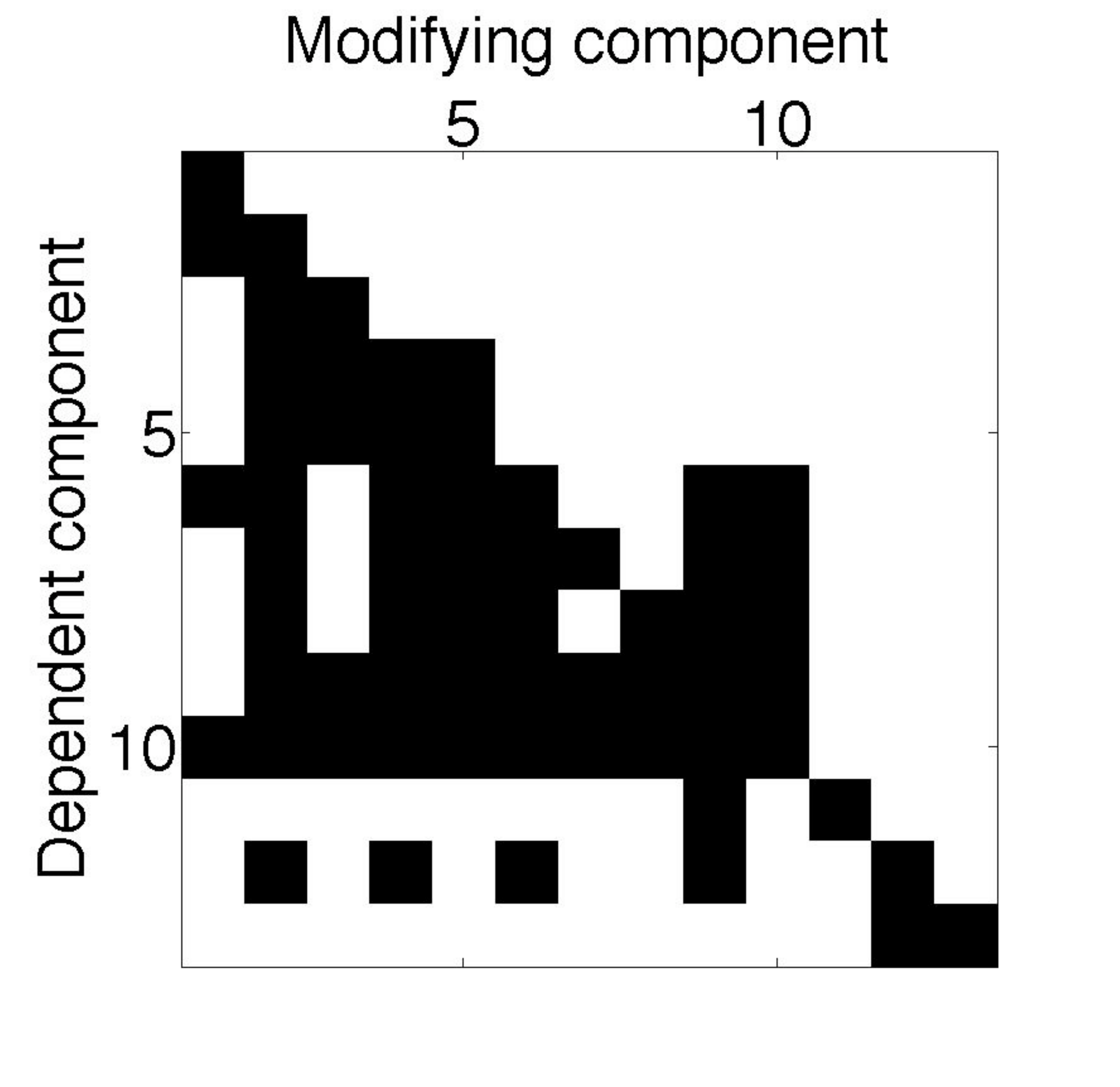}
\caption{Example dependency structure matrices (DSMs) with $n=13$ components.
  Black squares represent links. The DSM on the left was randomly generated
  to have fixed out-degree for each component. The DSM on the right
  represents the design of an automobile brake system \cite{Rivkin07}.  All
  diagonal elements are present because a component always affects its own
  cost.}
\label{DSM_patterns}
\end{figure}

We begin by considering the simplest case of \emph{fixed out-degree}, or
equivalently constant {\it connectivity}, in which each component affects
exactly $d$ components: itself and $d-1$ others.  The dependencies between
components are chosen at random with equal probability.  Cost reductions are
realized through the following series of innovation attempts:
\begin{enumerate}
\item Pick a random component $i$.
\item Use the DSM to identify the set of dependent components $\mathcal{A}_i
  = \{j \}$ whose costs depend on $i$.
\item Determine a new cost $c'_j$ for each component $j \in \mathcal{A}_i$
  from a specified probability distribution $f$.
\item If the sum of the new costs, $C'_{i} = \sum_{j \in \mathcal{A}_i}
  c'_j$, is less than the current sum, $C_{i}$, then each $c_j$
  is changed to $c'_j$. Otherwise, the new cost set is rejected.
\end{enumerate}

The costs are defined on $[0,1]$.  We assume a probability density function that for small values of $c_i$ has the form $f(c_i) \propto c_i^{\gamma-1}$, i.e. the cumulative distribution $F(c_i) = \int_0^{c_i} f(c) dc \propto c_i^\gamma$.  The exponent $\gamma$ specifies the difficulty of reducing costs of individual components, with higher $\gamma$ corresponding to higher difficulty.  This functional form is fairly general in that it covers any distribution with a valid power-series expansion at $c = 0$.

\section{Independent Components}
Let us first consider the process with a single component, as originally
studied by Muth \cite{Muth86}.  The generalization to $n$ independent
components is trivial, and only affects the constant of proportionality.
Letting $t$ represent the number of innovation steps, the cost at time $t$ is
equivalent to the minimum of $t$ independent, identically-distributed random
variables.  In the Supporting Information we show that the exact solution
when $\gamma = 1$ is
\begin{align}
c(t) = \frac{1}{1+t}.
\end{align}

We now present an intuitive derivation that gives the right scaling but underestimates the amplitude. At each innovation attempt a new cost $c'$ is drawn uniformly from $[0,1]$, and a successful reduction occurs if $c'$ is less than the current cost $c$. Since the distribution of new costs is uniform on $[0,1]$ the probability $\text{Prob}(c'< c)$ that $c'$ represents a reduction simply equals $c$. When a reduction does occur, the average value of $c'$ equals $c/2$.  In continuous time the rate of change of the average component cost is
\begin{align}\label{cdot}
\frac{dc}{dt} \sim -\left( \frac{c}{2} \right) \times \text{Prob}(c'<c) = -\frac{1}{2} c^2.
\end{align}
The solution to Eq. \eqref{cdot} gives the correct scaling of $c(t) \sim 1/t$ as $t \rightarrow \infty$.
One way to view this result is that cost reductions are proportional to the cost itself, leading to an exponential decrease in cost with each reduction event; however, each reduction takes exponentially longer to achieve as the cost decreases.  The competition between these two exponentials yields a power law. Muth's result is an application of extreme value theory, in which the mean value of the minimum of $m$ independent and identically distributed random variables evolves as a power law in $m$ when the variables are drawn from a distribution with a power-series expansion around zero.

\section{Interacting Components, Fixed Out-Degree}
Now consider an $n$-component process with fixed out-degree, where each
component affects exactly $d-1$ other, dependent components.  Following steps
similar to those that led to Eq.~\eqref{cdot}, let $c(t)$ denote the current
total cost of all components.  The average change in $c$ due to the next
improvement is given by summing over all changes that result in a new cost
$c'<c$, and multiplying the changes by their respective probabilities:
\begin{align*}
 \left< \Delta c \right> \approx -\int_0^c p(c') (c-c') \; dc'.
\end{align*}
This integral is performed in the Supporting Information.  The average change per timestep is
\begin{align}\label{approximate_diffeq}
\frac{dc}{dt} \approx -K c^{\gamma d + 1},
\end{align}
where
\begin{align*}
K = \frac{1}{n} d^{\gamma d + 1} \gamma^d \prod_{j=0}^{d-1} B(\gamma,j\gamma + 2),
\end{align*}
and $B(p,q)$ is the Beta function. The solution to Eq.~\eqref{approximate_diffeq} is
\begin{align}
\label{approximate_mean}
c(t) = c(0) \left(\frac{t}{t_0} + 1\right)^{-\frac{1}{\gamma d}},
\end{align}
with $t_0 = [K c(0)^{\gamma d} \gamma d]^{-1}$.  

\begin{figure}[t]
\noindent \includegraphics[width = .5\textwidth]{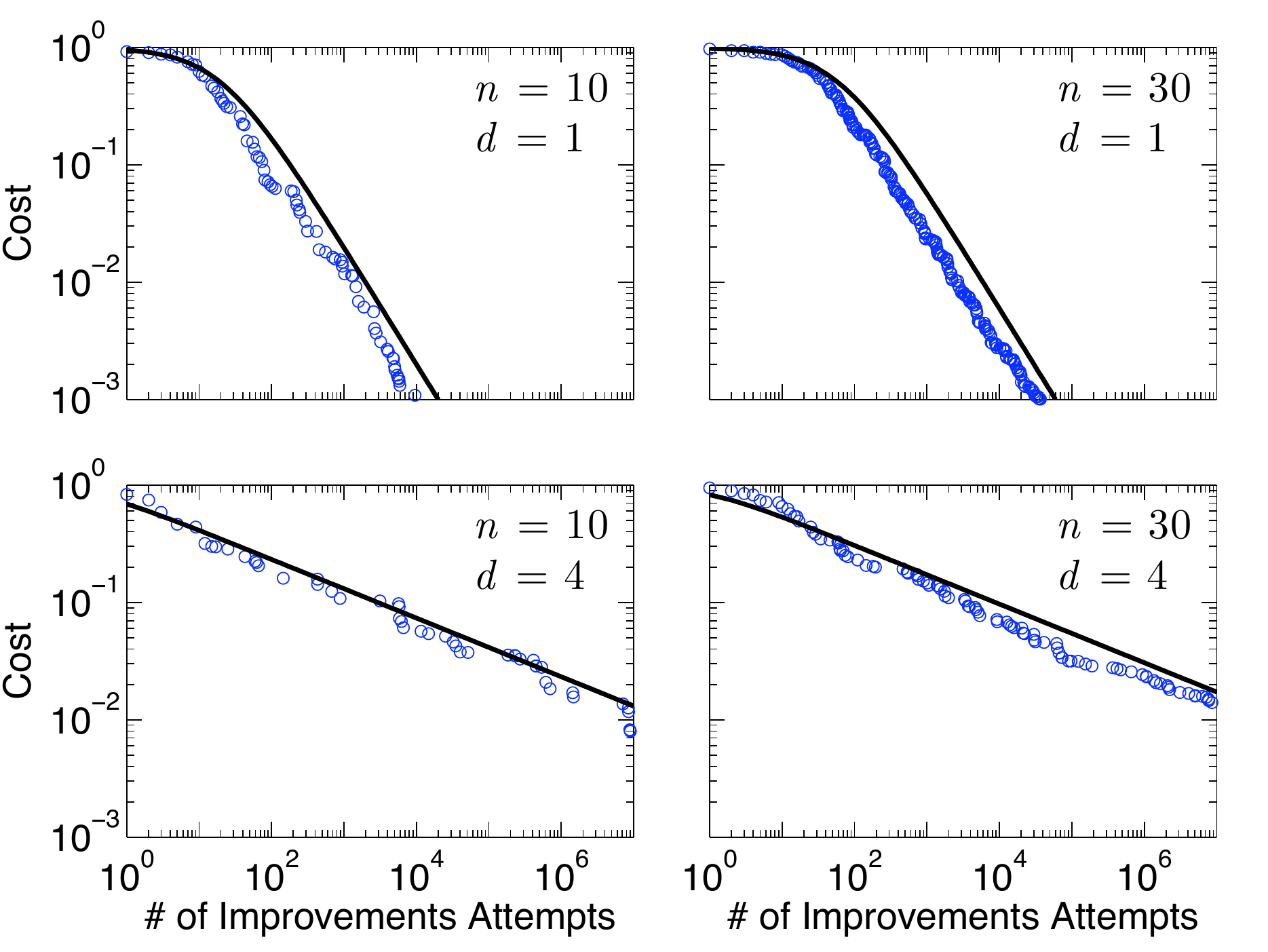}
\caption{Simulations of single realizations of the model for various numbers
  of components $n$ and design complexity $d^* = d$, using DSMs with constant
  $d$. The plots are for $n=10,30$ and $d=1,4$. The solid curves represent
  Eq.~\eqref{approximate_mean}.}
\label{single_runs}
\end{figure}

We compare our prediction in Eq.~\eqref{approximate_mean} to simulations in
Fig. 3.  Initially each component cost $c_i$ is set to $1/n$,
so that the total initial cost $c(0)=1$, and we arbitrarily choose
$\gamma=1$.  Our theory correctly predicts the asymptotic power law scaling
of the simulated performance curves.  Furthermore our theory predicts an
initial downward concavity that lasts for a time $t_0$; this prediction is
accurate for small $d$ (but for large $d$ it underestimates the concavity).
As given by Eq.~\eqref{approximate_mean}, increasing the number of components
$n$ extends the duration of the initial concavity.

The above asymptotic solution for $c(t)$ is consistently somewhat larger than the average value in the simulation.  This discrepancy stems from the approximations involved in deriving Eq.~\eqref{approximate_mean}.  In the Supporting Information the average cost is derived by other methods that compute the distribution.  These calculations yield the correct amplitude but only apply in the limit $t \rightarrow \infty$.

The salient result of this section is that the exponent $\alpha = 1/(\gamma d)$ of the performance curve is directly and simply related to the out-degree $d$, which can be viewed as a measure of the complexity of the design, and $\gamma$, which characterizes the difficulty of improving individual components in the limit as the cost goes to zero.  If $\gamma d =1$ then $\alpha = 1$ and the progress ratio $2^{-1/(\gamma d)}$is $50\%$.  If $\gamma d =3$ then $\alpha = 1/3$ and the progress ratio is approximately $80\%$, a common value observed in empirical performance curves.

\section{Interacting Components, Variable Out-Degree}

When the out-degree of each component is variable the situation is more
interesting and more realistic because components may differ in their rate of
improvement \cite{Rivkin07}.  Slowly improving components can create
bottlenecks that hinder the overall rate of improvement.  To understand why
such bottlenecks occur it is important to realize that the rate of
improvement of a given component depends on all the clusters $A_j$ of which
it is a member.  (Recall that $A_j$ is the set of components affected by
$j$.)  As illustrated in Fig. 4, there are two ways to reduce
the cost of component $i$:
\begin{enumerate}
\item Pick $i$ and improve cluster $\mathcal{A}_i$.
\item Pick another modifying component $j$ that affects $i$ and improve
  cluster $\mathcal{A}_j$.
\end{enumerate}
As we shall now show, the limiting rate for the improvement of a given
component $i$ is determined not only by the out-degree of the component
itself, but also by the properties of the clusters $A_j$ of the modifying
components that affect it.

First consider process (1) in which component $i$ is picked, corresponding to
the dotted ellipse in Fig. 4.  Let $C_i = \sum_{k \in
  \mathcal{A}_i} c_k$ be the sum of the costs of the $d_i$ dependent
components in $\mathcal{A}_i$.  Since $d_i$ new costs are drawn independently
at each time step the generation of new costs is equivalent to picking a
point with uniform probability in a $d_i$-dimensional hypercube.  The
combinations of component costs that reduce the total cost lie within the
simplex defined by $\sum_{k \in \mathcal{A}_i} c'_k < C_i$, where $c'_k$ are
the new costs.  The probability of reducing the cost is therefore the ratio
of the simplex volume to the hypercube volume,
\begin{align}
p_i = \frac{(C_i^{d_i}/d_i!)}{(1/n)^{d_i}} = \frac{(nC_i)^{d_i}}{d_i!},
\end{align}
which is a decreasing function of $d_i$.  Thus a component with a higher
out-degree (greater connectivity) is less likely to be improved when chosen.
(This is essentially the reason why $\alpha$ decreases with $d$ when the
out-degree is fixed.)

Let us now consider case (2) in which component $i$ improves when another
modifying component $j$ that affects it is chosen (lying inside the dashed
ellipse in Fig. 4).  Any component $j$ whose cluster contains
$i$ can cause an improvement.  Let $d_j^i$ be the out-degree of the modifying
components $j$ that affect $i$.  Then the overall rate at which $i$ improves
is determined by $d_i^{\rm min} = \min_j \{d_j^i \}$, i.e. by the component
$j$ most likely to cause an improvement in $i$.  The overall improvement rate
for the whole technology is then determined by the slowest improving
component $i$.  Thus the design complexity is more generally given by
\begin{equation}
d^* = \max_i \{d_i^{\rm min}\}.
\label{designComplexity} 
\end{equation}
Note that in the case of constant out-degree $d$ this reduces to $d^* = d$.  In every case we have studied $\alpha = 1/(d^* \gamma)$ correctly predicts the mean rate of improvement when $t$ is sufficiently large (see Fig. 5). 

\section{Fluctuations}
\begin{figure}[t]
  \center
\noindent
{\includegraphics[width=0.275\textwidth]{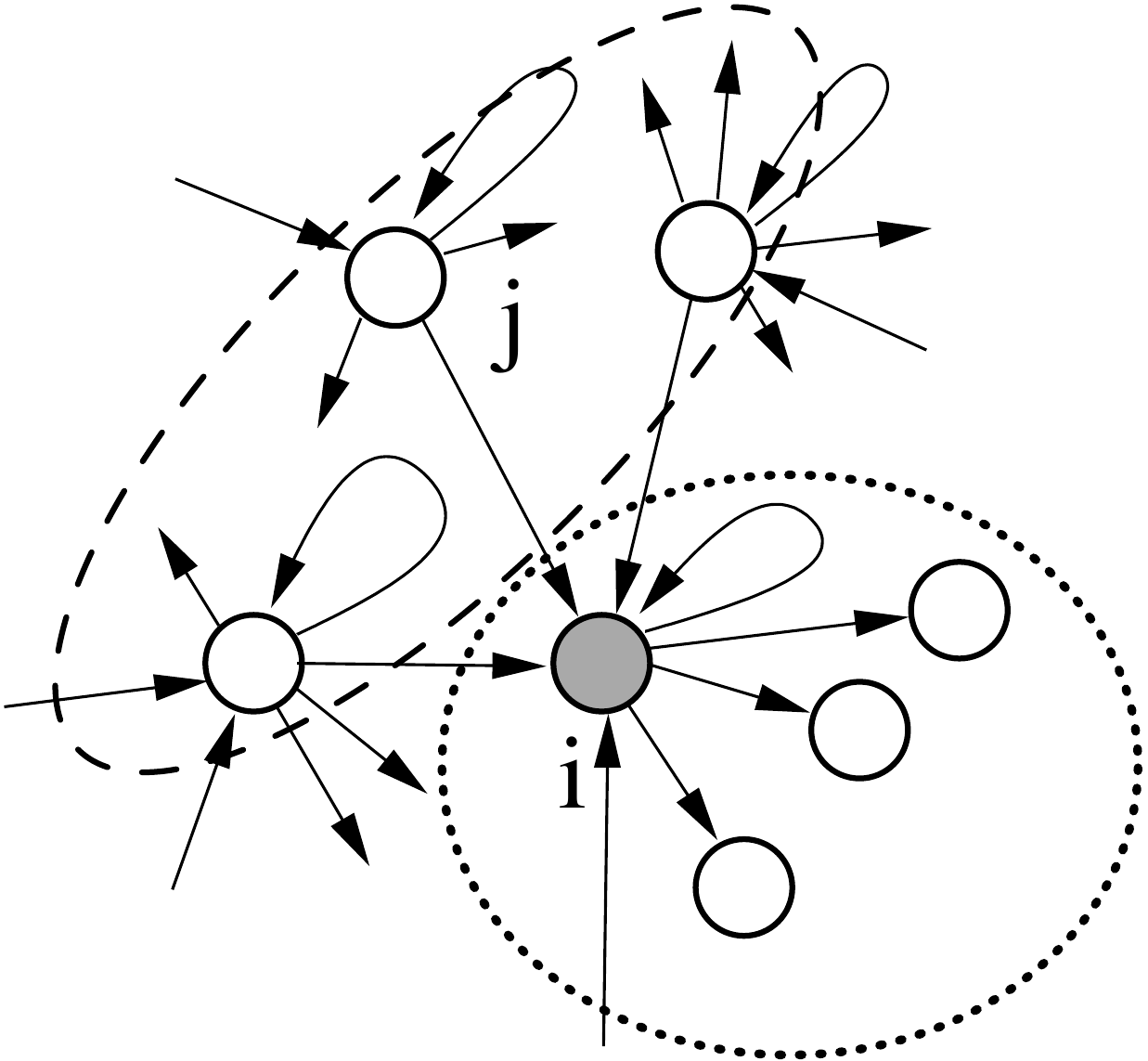}}
\caption{A component $i$ (shaded circle), together with the depedent
  components $\mathcal{A}_i$ that are affected by $i$ (dotted ellipse) and
  the modifying components that affect $i$ (dashed ellipse).  The arrow from
  $j$ to $i$ indicates that a change in cost of component $j$ affects the
  cost of $i$.}
\label{nodepair}
\end{figure}

The analysis we have given provides insight not only into the mean behavior, but also into the fluctuations about the mean, which can behave quite differently depending on the properties of the DSM.  In Fig. 5 we plot two individual trajectories of cost vs. time for each of three different DSMs.  The trajectories fluctuate in every case, but the amplitude of fluctuations is highly variable.  In the left panel  the amplitude of the fluctuations remains relatively small and is roughly constant in time when plotted on double logarithmic scale (indicating that the amplitude of the fluctuations is always proportional to the mean).  For the middle and right panels, in contrast, the individual trajectories show a random staircase behavior, and the amplitude of the fluctuations continues to grow for a longer time.  
\begin{figure*}[t]
\noindent
\parbox{.33\textwidth}{\center \includegraphics[width = .03\textwidth]{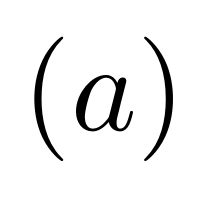}}\parbox{.33\textwidth}{\center \includegraphics[width = .03\textwidth]{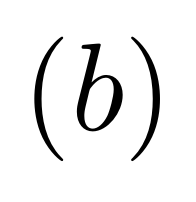}}\parbox{.33\textwidth}{\center \includegraphics[width = .03\textwidth]{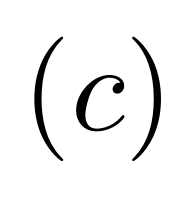}}

\noindent
\includegraphics[width = .33\textwidth]{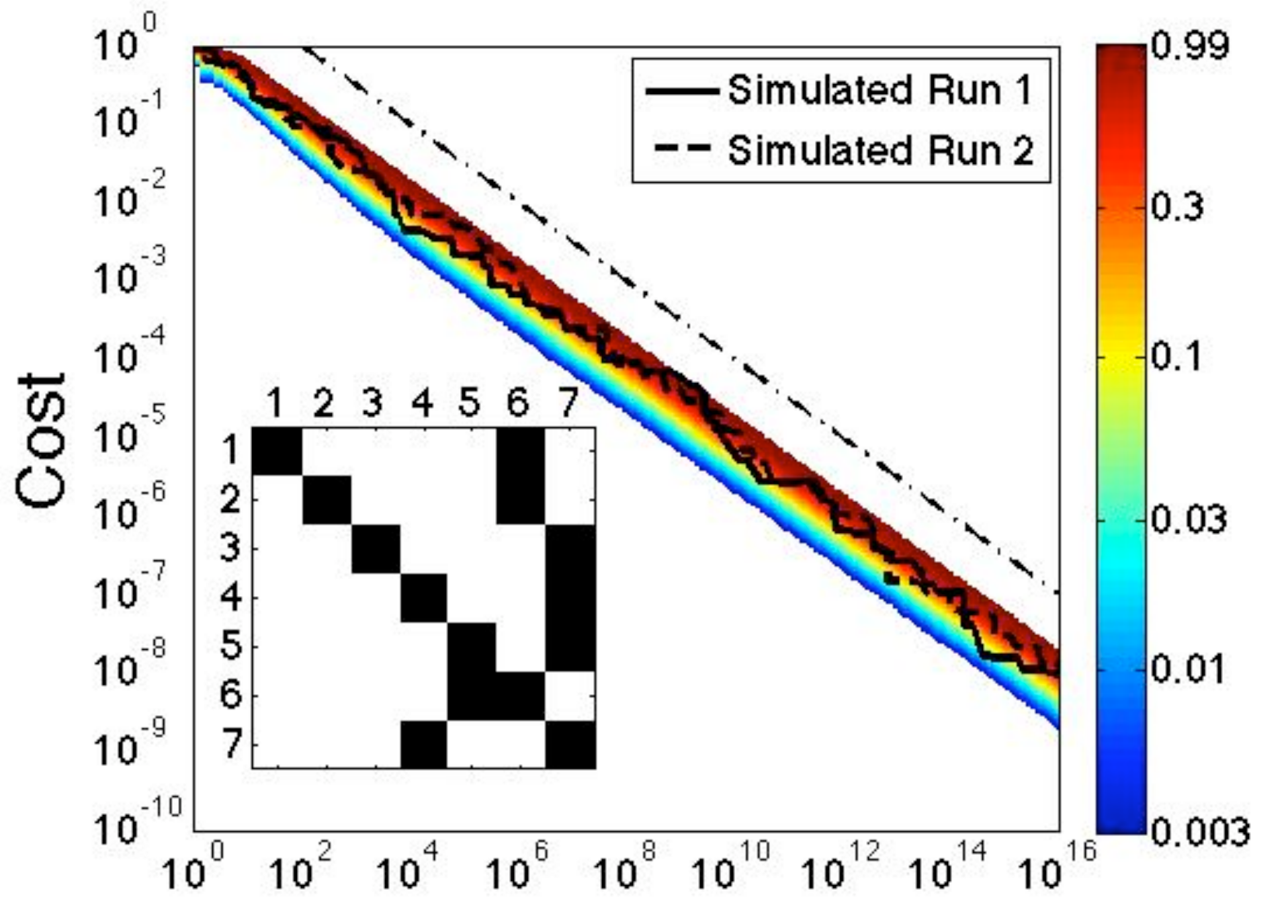}\includegraphics[width = .33\textwidth]{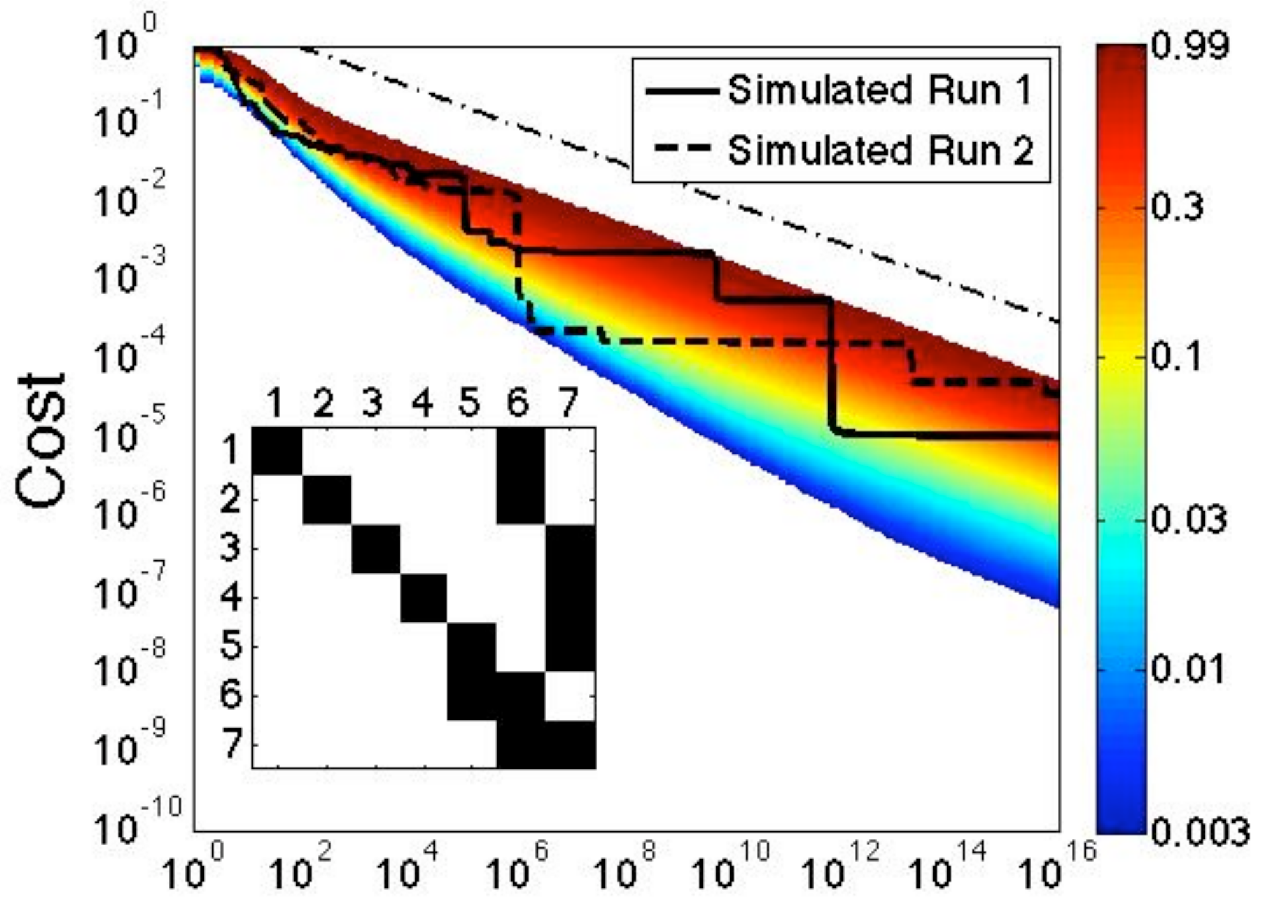}\includegraphics[width = .33\textwidth]{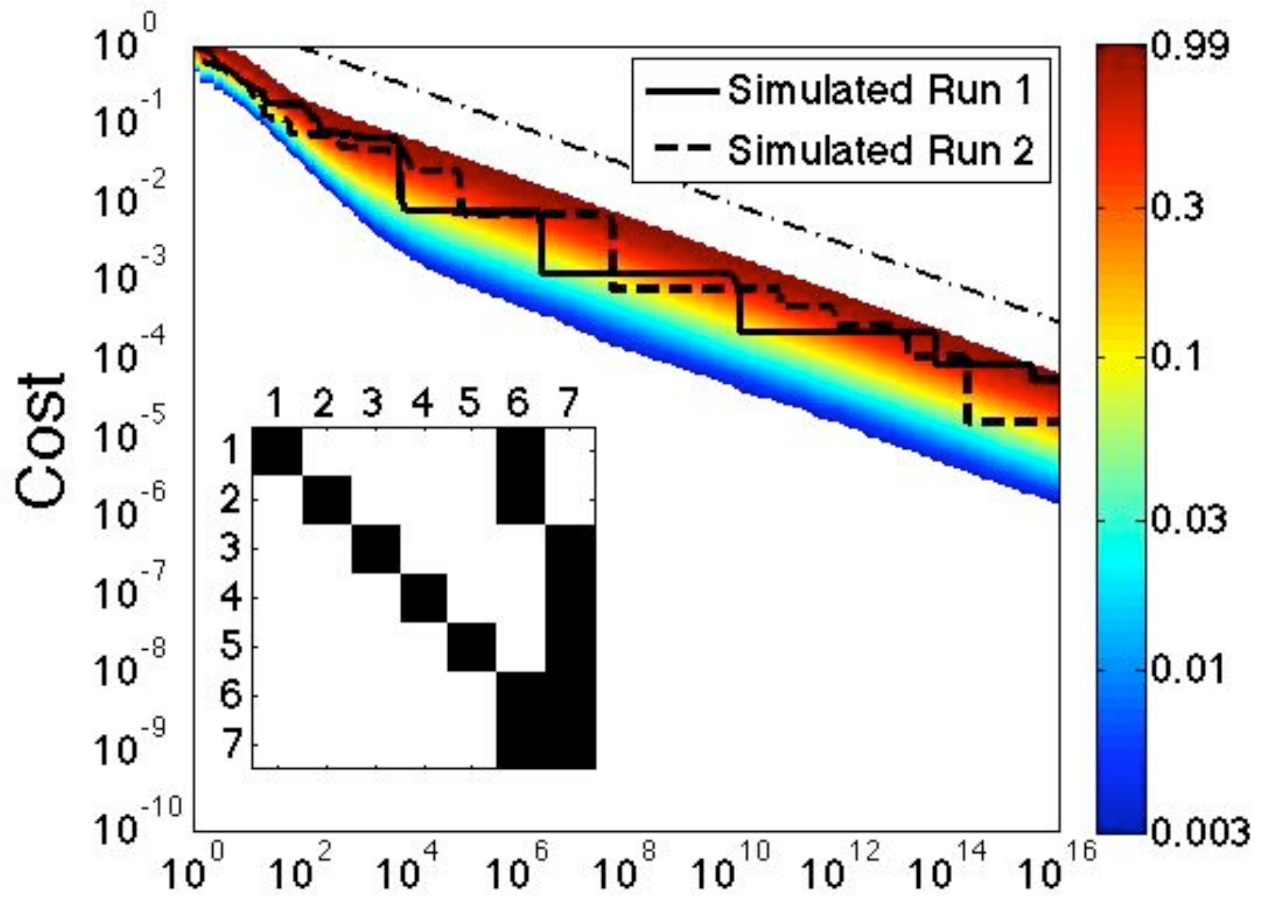}

\noindent
\includegraphics[width = .33\textwidth]{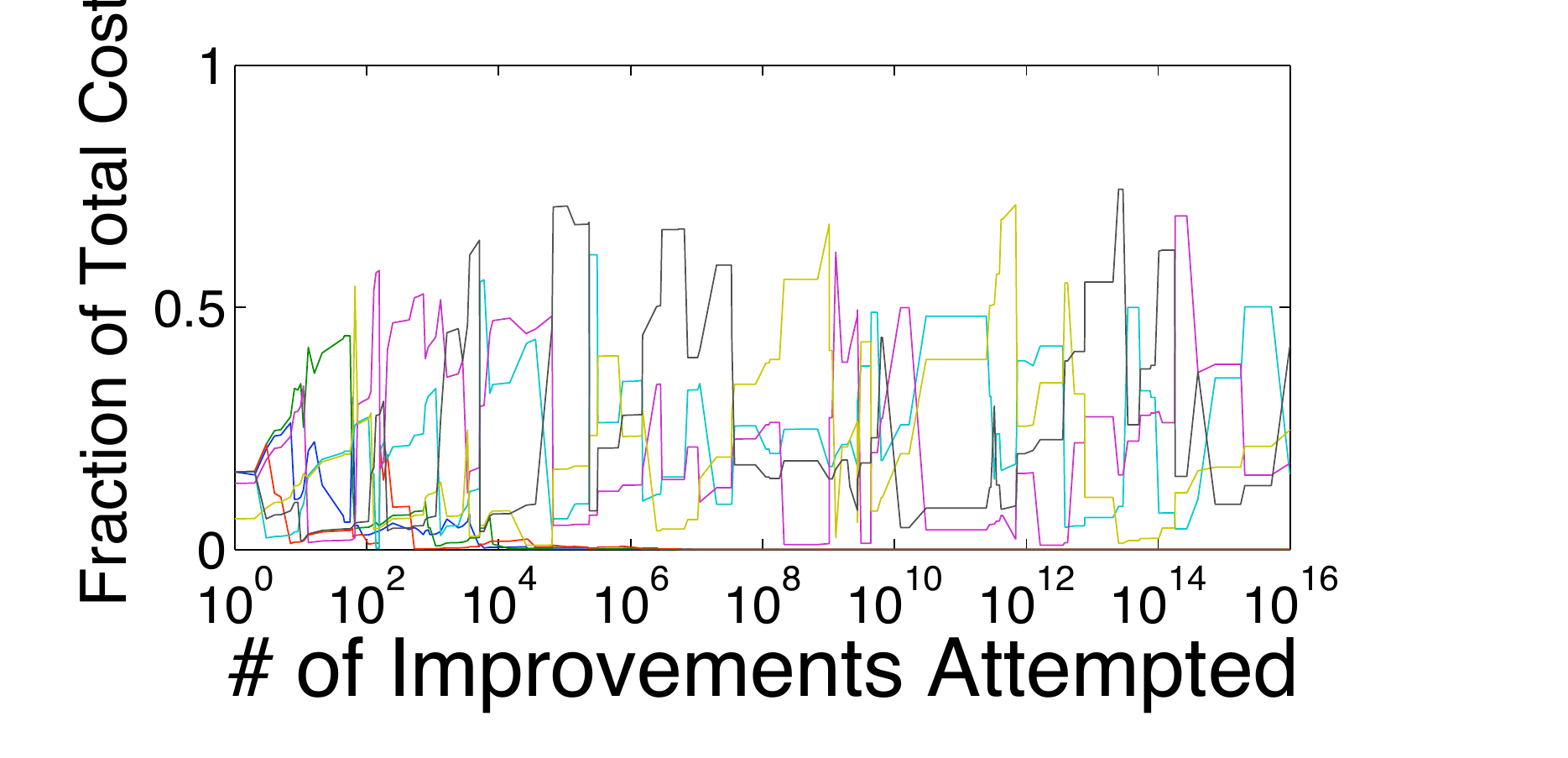}\includegraphics[width = .33\textwidth]{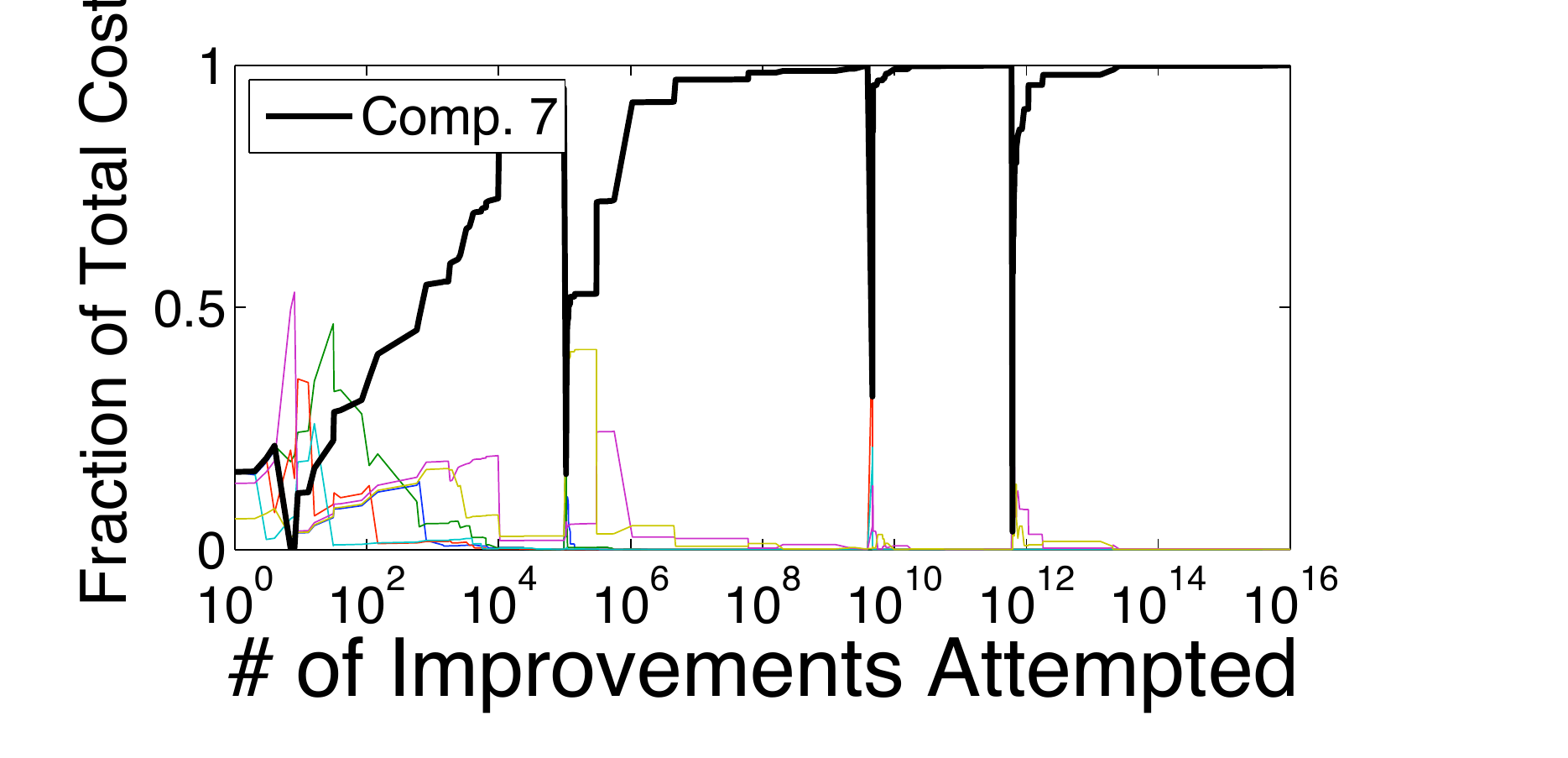}\includegraphics[width = .33\textwidth]{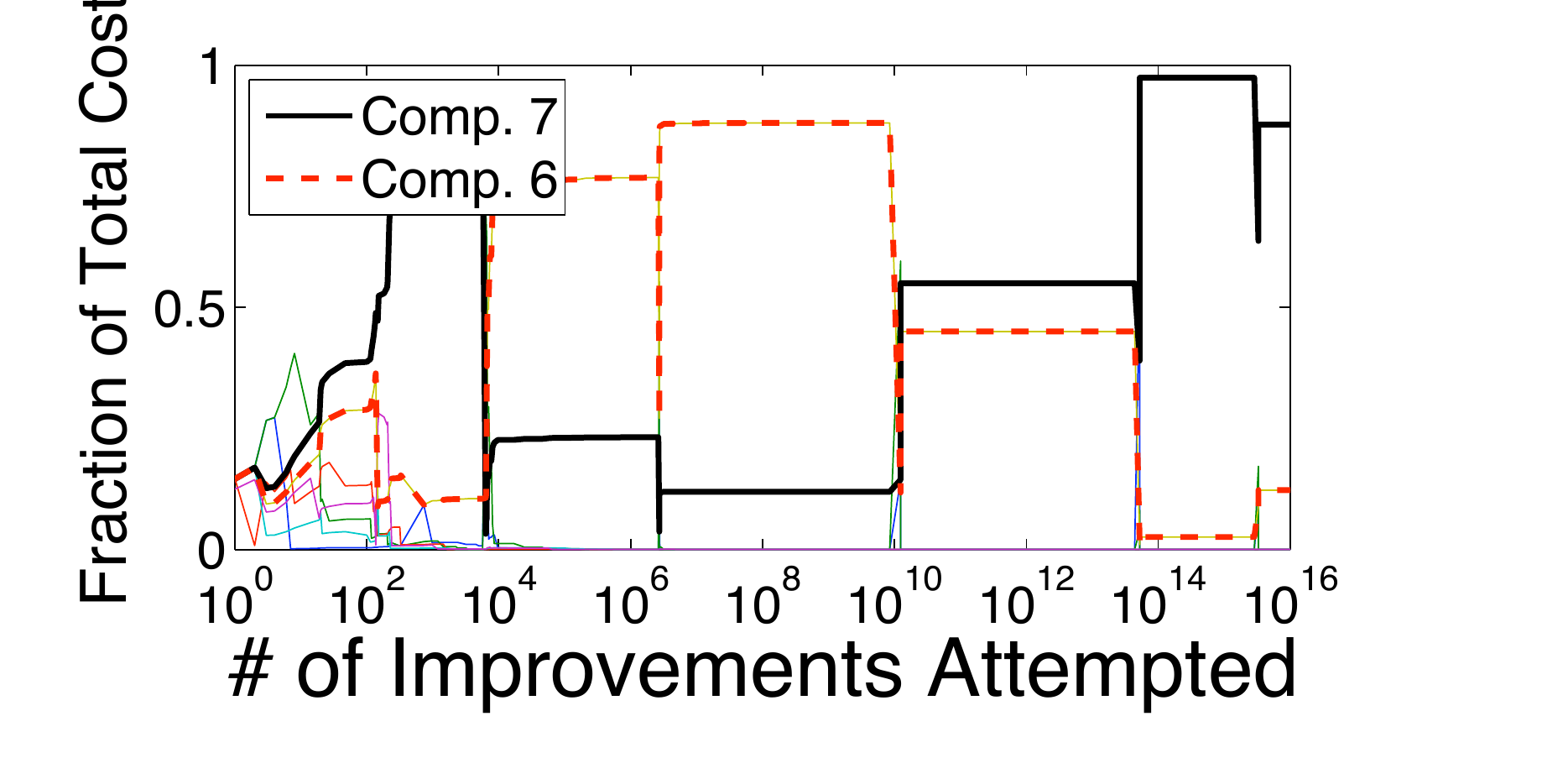}

\vspace{8pt}

\noindent
\parbox{.33\textwidth}{\includegraphics[width = .3\textwidth]{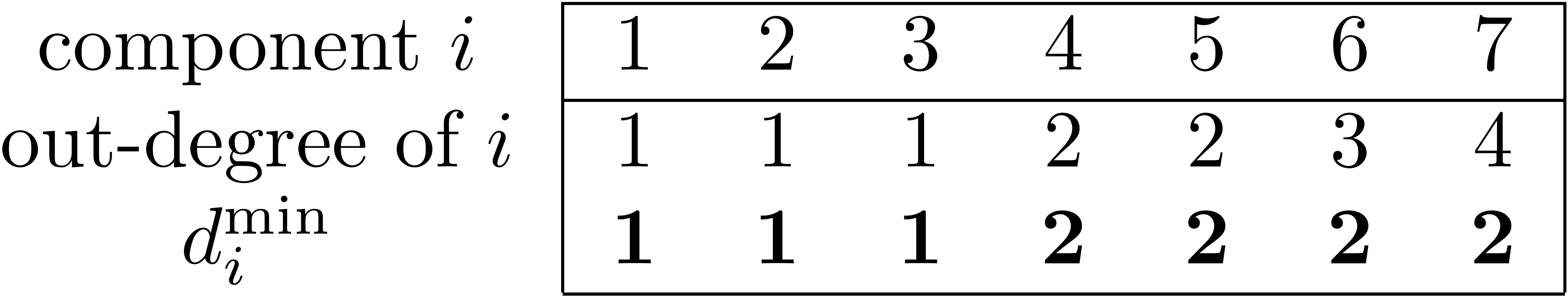}}\parbox{.33\textwidth}{\includegraphics[width = .3\textwidth]{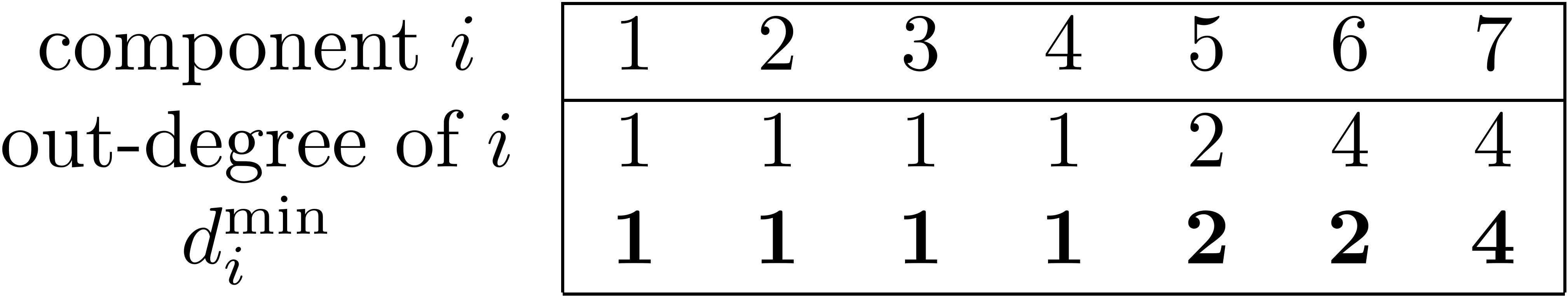}}\parbox{.33\textwidth}{\includegraphics[width = .3\textwidth]{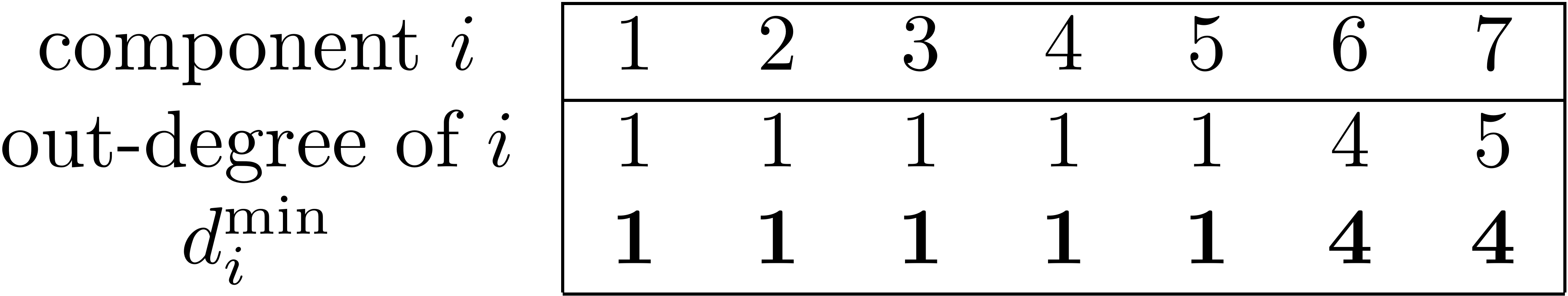}}
\caption{Evolution of the distribution of costs. Each figure in the top row shows a simulated distribution of costs as a function of time using the design structure matrix (DSM) in the lower left corner of each plot.  The upper dash-dot lines provides a reference with the predicted slope $\alpha = 1/(\gamma d^*$), with $\gamma = 1$; from left-to-right the slopes are $-1/2$, $-1/4$, and $-1/4$. The data  for each DSM are the result of 50,000 realizations, corresponding to different random number seeds.  The distributions are color coded to correspond to constant quantiles, i.e. the fraction of costs less than a given value at a given time. The solid and dashed black curves inside the colored regions represent two sample trajectories of the total cost as a function of time.  The DSMs are constructed so that in each case component 1 has the lowest out-degree and component 7 has the highest out-degree. Below each distribution we plot the fraction of the total cost contributed by each of the 7 components at any given time (corresponding to the first simulation run).  The components in Figs. (b) and (c) with the biggest contribution to the cost in the limit $t \rightarrow \infty$ are highlighted. The box at the bottom gives the value of $d_i^{min}$ for each component of the design.}
\label{distributions}
\end{figure*}

This behavior can be explained in terms of the improvement rates $d_i^{\rm
  min}$ for each component.  The maximum value of $d_i^{\rm min}$ determines
the slowest-improving components.  In the left panel the maximum value of
$d_i^{\rm min} = 2$.  This value is repeated four times.  After a long time
these four components dominate the overall cost.  However, since they have
the same values of $d_i^{\rm min}$ their contributions remain comparable, and
the total cost is averaged over all four of them, keeping the fluctuations
relatively small.  (See the lower panels of Fig. 5.)

In contrast, in the middle panel we illustrate a DSM where the
slowest-improving component (number $7$) has $d_i^{\rm min} = 4$ and the next
slowest-improving component (number $6$) has $d_i^{\rm min} = 2$.  With the
passage of time component $7$ comes to dominate the cost.  This component is
slow to improve because it is rarely chosen for improvement.  But in the rare
cases that component $7$ is chosen the improvements can be dramatic,
generating large downward steps in its trajectory.  The right case
illustrates an intermediate situation where two components are dominant.

Another striking feature of the distribution of trajectories is the
difference between the top and bottom envelopes of the plot of the
distribution vs. time.  In every case the top envelope follows a straight
line throughout most of the time range.  The behavior of the bottom envelope
is more complicated; in many cases, such as the left panel of
Fig. 5, it also follows a straight line, but in others, for
example the middle panel, the bottom envelope changes slope over a large time
range.  A more precise statement can be made by following the contour
corresponding to a given quantile through time.  All quantiles eventually
approach a line with slope $1/d^*$.  However, the upper quantiles converge to
this line quickly, whereas in some cases the lower quantiles do so much
later.  This slower convergence stems from the difference in improvement
rates of different components.  Whenever there is a dramatic improvement in
the slowest-improving component (or components), there is a period where the
next slowest-improving component (or components) becomes important.  During
this time the lower $d_i^{\rm min}$ value of the second component temporarily
influences the rate of improvement.  After a long time the slowest-improving
component becomes more and more dominant, large updates become progressively
more rare, and the slope becomes constant.

The model therefore suggests that properties of the design determine whether a technology's improvement will be steady or erratic. Homogeneous designs (with constant out-degree) are more likely to show an inexorable trend of steady improvement.  Heterogeneous designs (with larger variability in out-degree) are more likely to improve in fits and starts.  High variability among individual trajectories can be interpreted as indicating that historical contingency plays an important role. By this we mean that the particular choice of random numbers, rather than the overall trend, dominates the behavior.  In this case progress appears to come about through a series of punctuated equlibria.

To summarize, in this section we have shown that the asymptotic magnitude of
the fluctuations is determined by the {\it number of critical bottlenecks},
defined as the number of components that have the maximum value of $d_i^{\rm
  min}$.  In the constant out-degree case all of the components are
equivalent, and this number is just $n$.  In the variable out-degree case,
however, this number depends on the details of the DSM, which influence
$d_i^{\rm min}$. The fluctuations decrease as the number of worst bottlenecks
increases.

\section{Testing the Model Predictions}

Our model makes the testable prediction that the rate of improvement of a technology depends on the design complexity, which can be determined from a design structure matrix. The use of DSMs to analyze designs is widespread in the systems engineering and management science literature.  Thus, one could potentially examine the DSMs of different technologies, compute their corresponding design complexities, and compare to the value of $\alpha$ based on the technology's history\footnote{One problem that must be considered is that of resolution.  As an approximation a DSM can be constructed the coarse level of entire systems, e.g. ``electrical system", ``fuel system", or it can be constructed more accurately at the microscopic level in terms of individual parts.  In general these will give different design complexities.}.

This test is complicated by the fact that $\alpha$ also depends on $\gamma$,
which describes the inherent difficulty of improving individual components,
which in turn depends on the inherent difficulty of the innovation problem as
well as the collective effectiveness of inventors in generating improvements.
The exponent $\gamma$ is problematic to measure independently.  Nonetheless,
one could examine a collection of different technologies and either assume
that $\gamma$ is constant or that the variations in $\gamma$ average
out. Subject to these caveats, the model then predicts that the design
complexity of the DSM should be inversely proportional to the estimated
$\alpha$ of the historical trajectory.  A byproduct of such a study is that
it would yield an estimate of $\gamma$ in different technologies.

To compare the model predictions to real data one must relate the number of attempted improvements to something measurable. It is not straightforward to measure the effort that has gone into improving a technology, and to compare to real data one must use proxies.  The most commonly used proxy is cumulative production, but other possibilities include cumulative investment, installed capacity, R\&D expenditure, or even time.  The best proxy for innovation effort is a subject of serious debate in the literature \cite{Koh06, Koh08, Goddard82, Sinclair00, Nordhaus09}.

\section{Possible Extensions to the Model}
There are a variety of possible ways to extend the model to make it more realistic.  For example, the model currently assumes that the design network described by the DSM is constant through time, but often improvements to a technology come about by modifying the design network. One can potentially extend the model by adding an evolutionary model for the generation of new DSMs.

The possibility that the design complexity $d^*$ changes through time suggests another empirical prediction.  According to our theory, when $d^*$ changes,  $\alpha$ changes as well. One can conceivably examine an historical sequence of design matrices, compute their values of $d^*$, and compare the predicted $\alpha \sim 1/d^*$ to the corresponding observed values of $\alpha$ in the corresponding periods in time.  Our theory predicts that these should be positively correlated.

We have assumed a particular model of learning in which improvement attempts are made at random, with no regard to the history of previous improvements or knowledge of the technology.  An intelligent designer should be able to do better, drawing on his or her knowledge of science, engineering, and present and past designs.  While an accurate model of such a complex process is likely to be difficult, it is possible to alter the model here so that innovation attempts are relative to the current design, rather than independent of it.  This could potentially alter the predictions of the model.

\section{Discussion}
We have developed a simplified version of the model of Auerswald et
al. \cite{Auerswald00}, which predicts the improvement of cost as a function
of the number of innovation attempts.  While we have formulated the model in
terms of cost, one could equally well have used any performance measure of
the technology that has the property of being additive across the components.
Our analysis makes clear predictions about the trajectories of technology
improvement.  The mean behavior of the cost is described by a power law with
exponent $\alpha = 1/(d^* \gamma)$, where $d^*$ is the design complexity, and
$\gamma$ describes the intrinsic difficulty of improving individual
components.  In the case of constant connectivity (out-degree) the design
complexity is just the connectivity, but in general it can depend on details
of the design matrix, as spelled out in Eq.~\eqref{designComplexity}.  In
addition, the range of variation in technological improvement trajectories
depends on the number of critical bottlenecks.  This number coincides with
the total number of components $n$ in the case of constant connectivity, but
in general the number of worst bottlenecks is less than this, and depends on
the detailed arrangement of the interactions in the design.

Many studies in the past have discussed effects that contribute to technological improvement, such as learning-by-doing, R\&D, or capital investment.  Our approach here is generic in the sense that it could apply to any of these.  As long as these mechanisms cause innovation attempts that can be modeled as a process of trial and error, any of them can potentially be described by the model we have developed here.

Our analysis makes a new contribution by connecting the literature on the
historical analysis of performance curves to that on the engineering design
properties of a technology.  We make a prediction about how the features of a
design influence its rate of improvement, focusing attention on the
interactions of components as codified in the design structure matrix.
Perhaps most importantly, we pose several falsifiable propositions.  Our
analysis illustrates how the same evolutionary process can display either
historical contingency or steady change, depending on the design.  Our theory
suggests that it may be possible to influence the long-term rate of
improvement of a technology by reducing the connectivity between the
components.  Such an understanding of how the design features of a technology
affect its evolution will aid engineering design, as well as science and
technology policy.

\begin{acknowledgments}
  JM, DF, and JT gratefully acknowledge financial support from NSF Grant
  SBE0738187 and SR similarly acknowledges support from NSF Grant
  DMR0535503. We thank Yonathan Schwarzkopf, and Aaron
  Clauset for helpful conversations and suggestions. We thank Sidharth Rupani
  and Daniel Whitney for useful correspondence.
\end{acknowledgments}

\bibliographystyle{plain}
\bibliography{recipe_May30}
\end{article}
\end{document}